# The impact of local pinning sites in magnetic tunnel junctions with non-homogeneous free layers

*Alex. S. Jenkins, Leandro Martins, Luana Benetti, Alejandro Schulman, Pedro Anacleto, Marcel Claro, Elvira Paz, Ihsan Çaha, Francis Leonard Deepak, Ricardo Ferreira*

International Iberian Nanotechnology Laboratory, INL, Av. Mestre José Veiga s/n, 4715-330, Braga, Portugal.

**Pinning at local defects is a significant road block for the successful implementation of technological paradigms which rely on the dynamic properties of non-trivial magnetic textures. In this report a comprehensive study of the influence of local pinning sites for non-homogeneous magnetic layers integrated as the free layer of a magnetic tunnel junction is presented, both experimentally and with corresponding micromagnetic simulations. The pinning sites are found to be extremely detrimental to the frequency controllability of the devices, a key requirement for their use as synapses in a frequency multiplexed artificial neural networks. In addition to describing the impact of the local pinning sites in the more conventional NiFe, a vortex-based magnetic tunnel junction with an amorphous free layer is presented which shows significantly improved frequency selectivity, marking a clear direction for the design of future low power devices.**

Defect-induced pinning has long been perceived as a significant challenge for the real-world implementation of many potential applications which aim to harness the dynamical behaviour of complex magnetic textures, the prime example being the so-called 'racetrack memory' [1,2], where local pinning strongly affects the mobility of the magnetic texture under investigation [3]. Whilst significant study has focussed on the mobility of domain walls and skyrmions in nanowire configurations, there remains much less understanding of the effect of pinning in magnetic tunnel junctions with free layers that have a non-homogenous magnetisation. Vortex-based magnetic tunnel junctions (vMTJ) can be simplistically considered as comprising of three layers, a fixed magnetic layer whose magnetisation does not change, an insulating tunnelling layer, and a free magnetic layer, whose ground state is the magnetic vortex, where the in-plane magnetisation forms decreasingly small loops, before becoming out-of-plane at the central vortex core.

Significant interest has been shown recently in vortex-based frequency selective rectifiers [4,5], where individual nano-devices operate as tunable rf to dc transducers, with each device operating at a specific frequency, allowing for a frequency multiplexing approach which can be utilised in designing complex neuromorphic architectures [6,7]. The key element in such devices is the ability to precisely control the frequency of operation of each of these nano-rectifiers, which is usually achieved by controlling the geometry of the individual nanopillars. The magnetisation dynamics of magnetic vortices have been shown optically to be strongly dependent on the pinning sites created by structural defects in polycrystalline NiFe nano-disks [8–14], however there remains limited electrical characterisation of vortex-based magnetic tunnel junctions.

In this work, we show how the energy landscape induced by defects in polycrystalline magnetic materials causes the magnetic vortex being pinned, resulting in a clear sub-threshold pinned dynamic behaviour as well as the super-threshold conventional gyrotropic motion which is more commonly presented in the literature. Furthermore, we show how the sub-threshold dynamics are controlled, not by the device geometry, but by the local energy landscape of the pinning site, which is extremely detrimental to low-power operation of the devices as nano-rectifiers. One possible solution to the local pinning sites is by utilising a soft amorphous magnetic material, for example CoFeSiB, which is integrated as the free layer of a magnetic tunnel junction with a significantly reduced impact of local pinning sites.

**Results**

The devices under investigation are magnetic tunnel junctions with a CoFe(2.0)/Ru(0.7)/CoFeB(2.6nm) synthetic antiferromagnet (SAF) acting as the pinned layer and a CoFeB(2.0)/Ta(0.2)/**X** composite layer as the free layer separated by a thin MgO(1.0nm) layer. The soft magnetic material, **X**, included in the free layer composite is NiFe(7.0nm) similar to those discussed in detail in ref. [15], unless otherwise stated (i.e. Figure 4 and Figure 5). The patterned devices undergo an annealing treatment of 330°C for 2 hours in a 1 T magnetic field to align the axis of the pinning antiferromagnetic layer and crystallise the MgO. Similar behaviour was observed for devices with diameters ranging from d = 300-1000 nm.

**Analysing the local pinning through the spin-diode effect**

The spin-torque diode [16] effect is a well-established experimental technique for determining the resonant frequency of the dynamic modes in vMTJs [4,5,17], where applying a radio-frequency current results in a rectified voltage as the magnetisation response is transduced into an electrical signal via the tunnelling magnetoresistance response. In Figure 1 a) the magnitude of the rectified spin-diode voltage is presented as a function of the excitation frequency and power for a vMTJ.

The spin-diode presented in Figure 1 a) shows the presence of a clear threshold, $P_{rf}^t$ (illustrated in the figure with a white line), below which the dynamical behaviour of the magnetic vortex is very different, characterised by a relatively low voltage, higher frequency broad resonant response. Above the threshold value, a larger low frequency rectified voltage is observed. The in-plane magnetic field dependence of the dynamic response is plotted in Figure 1 b) and c), for excitation frequencies above and below $P_{rf}^t$, (i.e. $P_{rf}$ = 10 and 150 µW, respectively). The frequency of the sub-threshold mode can be seen to vary strongly as a function of the in-plane magnetic field, whereas the super-threshold behaviour has a more constant frequency response, at frequencies consistent with the gyrotropic motion of the vortex core published in the literature (e.g. refs. [5,18]) and predicted by the modified Thiele equation [19].

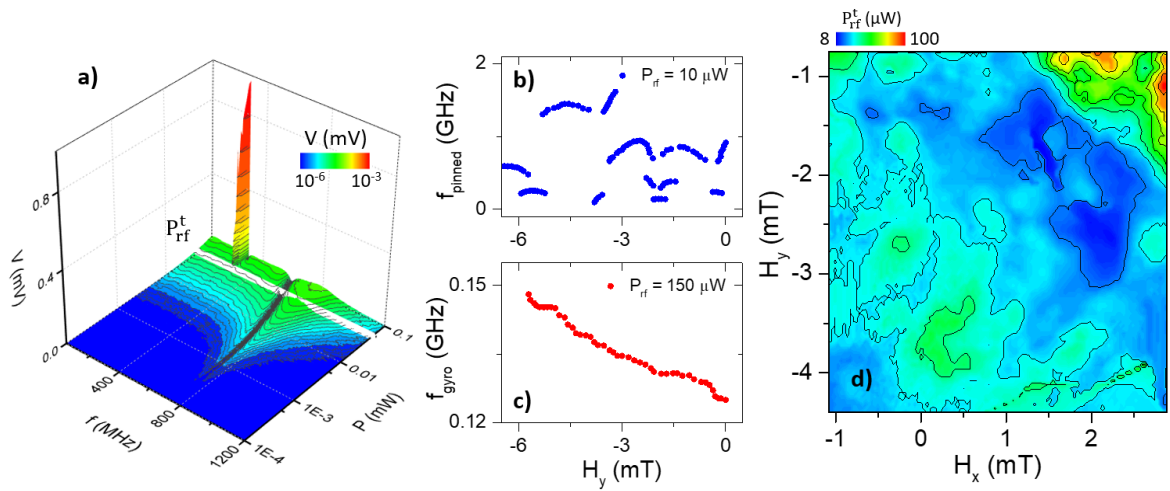

*Figure 1 – a) Magnitude of the experimentally determined spin-diode voltage as a function of the excitation frequency and power with the white line showing the threshold power, $P_{rf}^t$, at which the vortex core escape the pinning site and b), $P_{rf}^t$ as a function of in-plane magnetic field (Hx and Hy) and the resonant frequency of c) the pinned mode ($P_{rf}$ = 10 µW) and the gyrotropic mode ($P_{rf}$ = 150 µW). d) The threshold radio-frequency power, $P_{rf}^t$, as a function of the magnetic field applied along the x- and y-axis for an excitation current of $f_{gyro}$ = 115 MHz.*

In fact, the value of the threshold power, $P_{rf}^t$, can be relatively easily determined experimentally and is plotted in Figure 1 d) as a function of the in-plane field. A radio-frequency of f = 115 MHz is applied to the MTJ, corresponding to the gyrotropic mode of the vortex confined in a d = 500 nm nanopillar, and the power at which the measured rectified voltage exceeds a critical value (i.e. $V_{th}$ = 0.2 mV) is determined. The threshold power can be seen to vary significantly (from 8 to 100 µW), with the value depending strongly on the value of the in-plane magnetic field, and therefore the location of the vortex core.

**Micromagnetic simulations of the energy landscape**

Micromagnetic simulations can be utilised to better illustrate difference between the sub-threshold pinned dynamics and the super-threshold conventional gyrotropic mode. The micromagnetic simulations were performed using the mumax3 code for 500 nm nanopillars with typical NiFe material parameters ($M_{sat}$ = 740x10$^3$ A/m, $A_{ex}$ = 1.3x10$^{-11}$ J/m, alpha = 0.01). Although the pinning defects could be caused by other means (i.e. roughness [11]), in these simulations the local pinning defects were introduced by creating a random granular structure [14]. The nanopillar was separated into individual grains with an average size of 20 nm (values of the grain size in NiFe in the literature vary between 4 and 10 nm in ref. [14], 35 and 85 nm in ref. [8], 30 nm in ref [10]). The lower end of the range of grains sizes was selected as the free layer in this study is relatively thin ($t_{NiFe}$ = 7nm) and the grain size tends to increase with thickness [11]. Local pinning sites were created by randomly varying the saturation magnetisation of the grains between $M_{sat}$ and 0.85*$M_{sat}$ and the inter-region exchange at the boundaries between the grains was also reduced to 0.85*$A_{ex}$ at the granular interface.

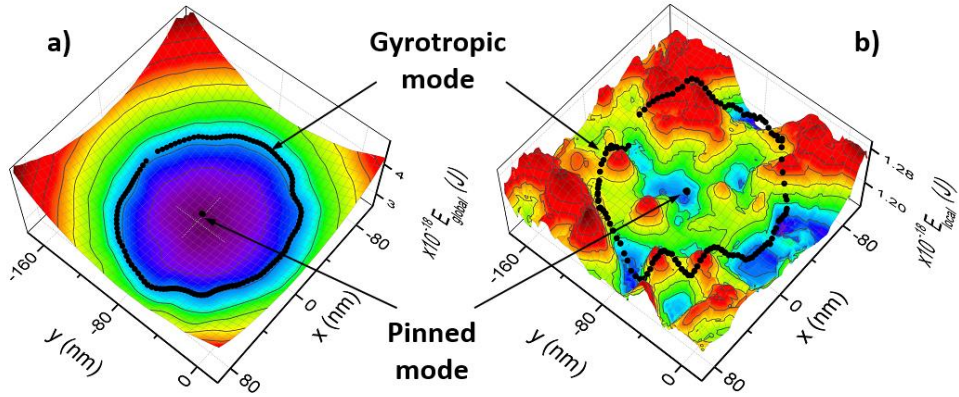

*Figure 2 – Energy landscape calculated with micromagnetic simulations considering a) the whole nanopillar and b) a 16 nm radius around the vortex core. The black dots shows typical trajectories of the sub-threshold pinned mode and the super-threshold gyrotropic mode.*

In Figure 2, the energy landscape has been plotted as a function of the x and y position within the nanopillar. The simulations were performed by initialising a magnetic vortex at different lateral positions (i.e., x and y) and waiting for a short time period (i.e. ~0.1 ns) for the core to stabilise and calculating the subsequent energy, which comprises of the demagnetising, exchange and Zeeman component energy terms. In Figure 2 a), the energy of the whole of the nanopillar is calculated, and results in a standard parabolic potential minima. Although the granular defects are very slightly visible, the majority of the energy landscape is dominated by the edges of the nanopillar, with the in-plane part of the magnetic vortex typically being > 500 times in size compared to the vortex core.

In Figure 2 b), however, the energy landscape is again calculated, except this time only the energy within an area comparable to the vortex core (i.e. a circle with radius r = 16 nm) is calculated, and the local energy minima felt by the vortex core is presented. The impact of the granular defects is now much more pronounced, with lots of local maxima and minima visible. To illustrate the sub-threshold and super-threshold dynamics, typical trajectories of the vortex core when strongly pinned and after escaping the pinning site (i.e. the gyrotropic mode) are plotted.

In order to validate the micromagnetic simulations with the experimental data in the sub-threshold regime, in Figure 3, the frequency of the pinned mode determined experimentally using the spin-diode effect ($P_{rf}$ = 10 µW) is compared with an FFT analysis of the magnetisation aligned along the easy-axis in the micromagnetic simulations, i.e. the $m_y$ component, as the magnetic vortex relaxes over 50 ns, presented as a function of the in-plane magnetic field (both transverse and along the MTJ easy axis, i.e. Hx and Hy respectively). There is good agreement between the experimentally acquired spin-diode effect and the micromagnetic simulations, with the resonant frequency varying strongly as a function of the in-plane magnetic field, with different regions of similar resonant frequency corresponding to different pinning sites.

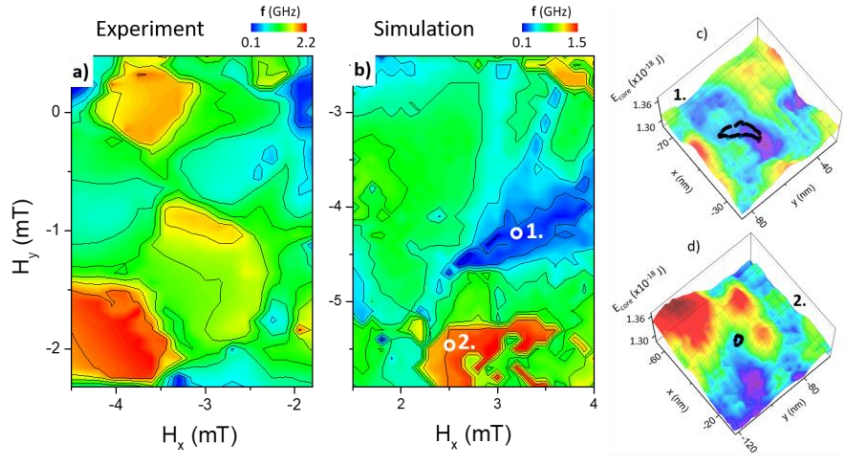

*Figure 3 – Resonant frequency of the sub-threshold pinned mode as determined a) experimentally via spin diode effect and via b) micromagnetic simulations, as a function of the in-plane magnetic field (Hx and Hy). Energy landscape as calculated with micromagnetic simulations for two different in-plane magnetic fields i.e. c) Hx = -4.2mT and Hy = 3.3 and d) Hx = -5.4 mT and Hy = 2.6 mT. The black line corresponds to the trajectory of the vortex core when excited with an rf signal of 1 mA and a frequency of 220 MHz.*

To further explore the different pinning sites and their consequent effect on the vortex dynamics, two magnetic fields were identified in Figure 3 b) (i.e [Hx, Hy] = -4.2, 3.3 mT and [Hx ,Hy] = -5.4,2.6 mT, labelled 1 and 2 respectively) with very different resonant frequencies ($f_1$ = 220 MHz and $f_2$ = 1400 MHz). These two fields correspond to a weakly pinned system (i.e. position 1) and a strongly pinned system (i.e. position 2). The energy landscapes calculated via micromagnetic simulations of these two different field values are presented in Figure 3 c) and d). Additionally, the typical trajectories of the sub-threshold pinned mode are presented for two resonant excitations ($f_1$ = 220MHz and $f_2$ = 1400 MHz for an excitation current of 1 mA). For position 1, the loosely pinned vortex core has a resonant frequency close to the global gyrotropic frequency, and the core is excited to relatively large orbits. This pinning site is spatially larger and is caused by two grains of relatively low saturation magnetisation being located next to one another (supplementary information Figure 1). In the case of position 2, the strongly pinned core can be resonantly excited by applying a frequency far from the global gyrotropic mode, and results in a relatively small orbit. This energy minima is caused by the intersection of multiple grain boundaries which result in a pronounced yet spatially localised minima (supplementary information Figure 2). The origin of the pinning site appears to have a pronounced impact on the vortex core dynamics leading to the possibility that analysis of the resonant frequency of the vortex core and subsequent comparison to micromagnetic simulations may provide an interesting tool for exploration of the granular structure of magnetic materials in the future.

**Reducing the influence of pinning with amorphous free layers**

Having identified the challenge posed by local pinning sites in vortex-based magnetic tunnel junctions, we now present two potential solutions for reducing the impact of these local energy minima on the vortex dynamics; removing granular defects by using an amorphous free layer and reducing the impact of the interface by increasing the free layer thickness. The work presented in the first half of this report uses a CoFeB/Ta/NiFe composite free layer, where the polycrystalline NiFe is chosen due to its magnetic softness, and the CoFeB is selected to ensure good texture in the insulating MgO barrier, and therefore subsequent high tunneling magnetoresistance. The CoFeB is amorphous as deposited, but crystalises into bcc (001) during the annealing which is an essential step to ensure high TMRs above 100%. The Ta layer is to ensure a separation of crystalinity between the NiFe and the CoFeB.

Whilst the CoFeB/Ta/NiFe composite layer has shown good results for large power excitations, the polycrstaline nature of the free layer results in additional pinning sites. In order to avoid this, an amorphous soft magnetic layer has been selected to replace the NiFe, namely CoFeSiB. $Co_{67}Fe_4Si_{14.5}B_{14.5}$ is an interesting candidate for magnetic tunnel junctions as it is amorphous after standard annealing processes [20] and has been previously investigated in micron-sized AlO-based magnetic tunnel junctions where lower switching fields were observed [21–23]. Figure 4 a) and b) shows the x-ray diffraction measurement of two thin films, NiFe and CoFeSiB respectively, after annealing at 330°C. A clear peak associated with fcc (111) is observed for the polycrystalline NiFe, whereas no peak is observed for the CoFeSiB.

As well as being amorphous in unpatterned thin films, in Figure 4 c) and d) the cross sectional transmission electron microscopy (TEM) is presented for two patterned magnetic tunnel junction devices, with c) CoFeB/Ta/NiFe and d) CoFeB/Ta/CoFeSiB free layers. By analysis of the Fast Fourier Transform (FFT), the NiFe image is shown to be consistent with a orthorhombic crystal with 111 zone axis, whilst the CoFeSiB appears amorphous after patterning and annealing.

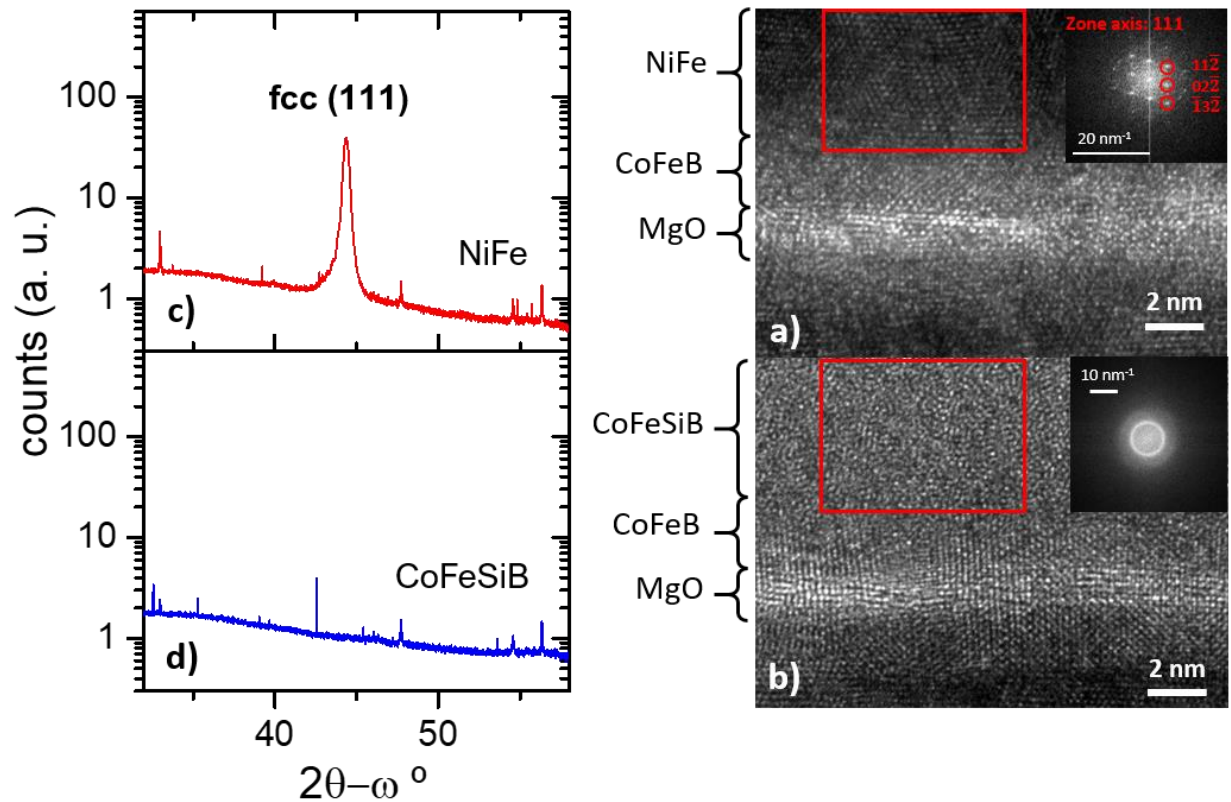

Figure 4 – x-ray diffraction pattern of a) NiFe and b) CoFeSiB annealed thin films as well as cross sectional TEM images of two patterned magnetic tunnel junction devices with c) CoFeB/Ta/NiFe and d) CoFeB/Ta/CoFeSiB free layers.

Having identified CoFeSiB as an amorphous free layer, even after annealing, nano-devices were patterned and their resonant frequency was investigated via the spin diode effect (similar to Figure 1 b) with an rf current of $P_{rf} = 1$ μW. As seen in Figure 5 a), the CoFeB(2.0)/Ta(0.5)/NiFe(7.0 nm) shows the sub-threshold behaviour discussed in Figure 1, with a drastic variation of the resonant frequency as a function of the in-plane magnetic field, as the core moves from one pinning site to another, each with markedly different energy landscapes and therefore resonant frequencies. When the NiFe is replaced with CoFeSiB(7 nm), shown in

Figure 5 b), the resonant frequency is seen to still vary significantly as a function of the in-plane magnetic field but with the impact being reduced by the amorphous nature of the CoFeSiB. Whilst there still exists a variation in frequency, this can be seen to be between 0.1 - 1 GHz (relative to the 0.1 – 2.5 GHz of the NiFe). The fact that the vortex continues to be in the sub-threshold pinned regime even with an amorphous free layer suggests that the impact of interfacial roughness and the polycrystalline CoFeB are still playing an important role in the energy landscape. In Figure 5 c), the thickness of the CoFeSiB is increased to 40 nm and the variation in resonant frequency can be seen to drastically reduce, with the spin diode response being observed mostly between 0.5 – 0.6 GHz, demonstrating a substantial reduction of frequency variation due to the local pinning sites.

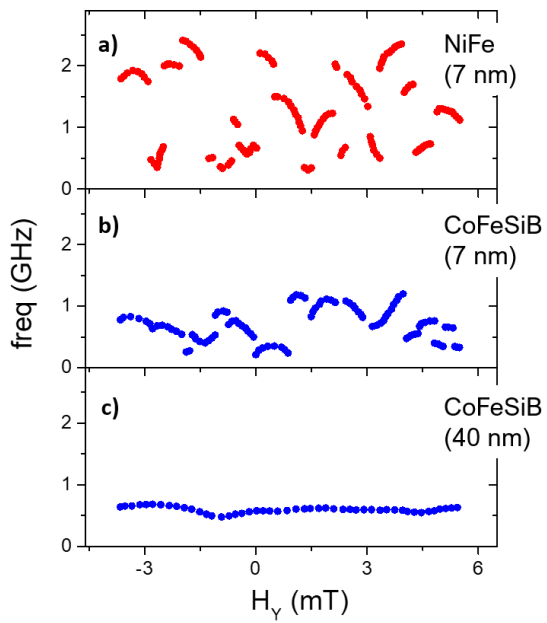

*Figure 5 – The measured resonant frequency as a function of the in-plane magnetic field for the magnetic tunnel junctions of 400 nm diameter, with a free layer comprised of a) CoFeB(2.0)/Ta(0.5)/NiFe(7 nm), b) CoFeB(2.0)/Ta(0.5)/CoFeSiB(7 nm) and c) CoFeB(2.0)/Ta(0.5)/CoFeSiB(40 nm)*

**Discussion**

In conclusion, the impact of local pinning sites is explored in vortex-based magnetic tunnel junctions, where a threshold behaviour is observed in the excitation power, below which the magnetic vortex remains pinned. In these pinned states, the resonant frequency of the nano-device is strongly impacted by the local energy landscape leading to a drastic variation in the resonant frequency. These pinning sites represent a significant challenge for the implementation of vortex-based frequency selective rectifiers, however by using an amorphous free layer, namely CoFeSiB, the impact of the vortex core pinning is significantly reduced. The integration of amorphous materials as the free layer of magnetic tunnel junctions with non-trivial magnetic textures offers a potentially 'pinning-free' system, which can dramatically improve device performance in the low-power regime.

## Acknowledgements

This work has received funding from the European Union's Horizon 2020 research and innovation programme under grant agreement No 101017098 (project RadioSpin), No 899559 (project SpinAge) and No 101070287 (project Swan-on-chip).

## Materials and Methods

Transmission electron microscopy (TEM) imaging of the samples were carried out with a double corrected FEI Titan G3 Cubed Themis equipped with a Super-X EDX detector, operated at 200 kV. The samples for TEM analysis were prepared by a dual-beam FEI Helios focused ion beam (FIB) following the standard lift-out procedure. In order to not destroy the formed oxide layer on the samples by Ga+ ions during the milling steps, two Pt layers were deposited: ∼200 nm by electron (e-) beam and ∼2 μm by ion (i-) beam. 30 kV ion beam was used for the bulk milling, and thereafter the lamella was thinned down to 100 nm in thickness. Finally, a 5 kV ion beam was used to remove the surface amorphization layer.

[1] S. S. P. Parkin, M. Hayashi, and L. Thomas, Magnetic domain-wall racetrack memory, Science (80-. ). **320**, 190 (2008).

[2] A. Fert, V. Cros, and J. Sampaio, Skyrmions on the track, Nat. Nanotechnol. 2013 83 **8**, 152 (2013).

[3] A. Thiaville, Y. Nakatani, J. Miltat, and Y. Suzuki, Micromagnetic understanding of current-driven domain wall motion in patterned nanowires, Europhys. Lett. **69**, 990 (2005).

[4] L. Martins, A. S. Jenkins, L. S. E. Alvarez, J. Borme, T. Böhnert, J. Ventura, P. P. Freitas, and R. Ferreira, Non-volatile artificial synapse based on a vortex nano-oscillator, Sci. Reports 2021 111 **11**, 1 (2021).

[5] D. Marković, N. Leroux, A. Mizrahi, J. Trastoy, V. Cros, P. Bortolotti, L. Martins, A.

Jenkins, R. Ferreira, and J. Grollier, Detection of the Microwave Emission from a Spin-Torque Oscillator by a Spin Diode, Phys. Rev. Appl. **13**, 044050 (2020).

[6] N. Leroux, A. Mizrahi, D. Marković marković, D. Sanz-Hernández, J. Trastoy, P. Bortolotti, L. Martins, A. Jenkins, R. Ferreira, and J. Grollier, Hardware realization of the multiply and accumulate operation on radio-frequency signals with magnetic tunnel junctions, Neuromorphic Comput. Eng. **1**, 011001 (2021).

[7] N. Leroux, D. Marković, E. Martin, T. Petrisor, D. Querlioz, A. Mizrahi, and J. Grollier, Radio-Frequency Multiply-and-Accumulate Operations with Spintronic Synapses, Phys. Rev. Appl. **15**, 034067 (2021).

[8] R. L. Compton and P. A. Crowell, Dynamics of a pinned magnetic vortex, Phys. Rev. Lett. **97**, 137202 (2006).

[9] R. L. Compton, T. Y. Chen, and P. A. Crowell, Magnetic vortex dynamics in the presence of pinning, Phys. Rev. B - Condens. Matter Mater. Phys. **81**, 144412 (2010).

[10] J. S. Kim, O. Boulle, S. Verstoep, L. Heyne, J. Rhensius, M. Kläui, L. J. Heyderman, F. Kronast, R. Mattheis, C. Ulysse, and G. Faini, Current-induced vortex dynamics and pinning potentials probed by homodyne detection, Phys. Rev. B - Condens. Matter Mater. Phys. **82**, 104427 (2010).

[11] T. Y. Chen, M. J. Erickson, P. A. Crowell, and C. Leighton, Surface roughness dominated pinning mechanism of magnetic vortices in soft ferromagnetic films, Phys. Rev. Lett. **109**, 097202 (2012).

[12] J. A. J. Burgess, A. E. Fraser, F. Fani Sani, D. Vick, B. D. Hauer, J. P. Davis, and M. R. Freeman, Quantitative magneto-mechanical detection and control of the Barkhausen effect, Science (80-. ). **339**, 1051 (2013).

[13] J. A. J. Burgess, J. E. Losby, and M. R. Freeman, An analytical model for vortex core pinning in a micromagnetic disk, J. Magn. Magn. Mater. **361**, 140 (2014).

[14] M. Möller, J. H. Gaida, and C. Ropers, Pinning and gyration dynamics of magnetic vortices revealed by correlative Lorentz and bright-field imaging, Phys. Rev. Res. **4**, 013027 (2022).

[15] A. S. Jenkins, L. San Emeterio Alvarez, R. Dutra, R. L. Sommer, P. P. Freitas, and R. Ferreira, Wideband High-Resolution Frequency-to-Resistance Converter Based on Nonhomogeneous Magnetic-State Transitions, Phys. Rev. Appl. **13**, 014046 (2020).

[16]  A. A. Tulapurkar, Y. Suzuki, A. Fukushima, H. Kubota, H. Maehara, K. Tsunekawa, D. D. Djayaprawira, N. Watanabe, and S. Yuasa, Spin-torque diode effect in magnetic tunnel junctions, Nature **438**, 339 (2005).

[17]  A. S. Jenkins, L. S. E. Alvarez, S. Memshawy, P. Bortolotti, V. Cros, P. P. Freitas, and R. Ferreira, Electrical characterisation of higher order spin wave modes in vortex-based magnetic tunnel junctions, Commun. Phys. 2021 41 **4**, 1 (2021).

[18]  S. Tsunegi, K. Yakushiji, A. Fukushima, S. Yuasa, and H. Kubota, Microwave emission power exceeding 10 μ W in spin torque vortex oscillator, Appl. Phys. Lett. **109**, 252402 (2016).

[19]  a Dussaux, B. Georges, J. Grollier, V. Cros, a V Khvalkovskiy, a Fukushima, M. Konoto, H. Kubota, K. Yakushiji, S. Yuasa, K. a Zvezdin, K. Ando, and a Fert, Large microwave generation from current-driven magnetic vortex oscillators in magnetic tunnel junctions., Nat. Commun. **1**, 8 (2010).

[20]  F. Celegato, M. Coïsson, E. Olivetti, P. Tiberto, and F. Vinai, A study of magnetic properties in CoFeSiB amorphous thin films submitted to furnace annealing, Phys. Status Solidi **205**, 1745 (2008).

[21]  Y. K. Kim, Magnetic tunnel junctions comprising amorphous NiFeSiB and CoFeSiB free layers, J. Magn. Magn. Mater. **304**, 79 (2006).

[22]  M. Löhndorf, S. Dokupil, J. Wecker, M. Rührig, and E. Quandt, Characterization of magnetic tunnel junctions (MTJ) with magnetostrictive free layer materials, J. Magn. Magn. Mater. **272**–**276**, 2023 (2004).

[23]  J. Y. Hwang, S. S. Kim, M. Y. Kim, J. R. Rhee, B. S. Chun, Y. K. Kim, T. W. Kim, H. B. Lee, and S. C. Yu, Switching characteristics of magnetic tunnel junction with amorphous CoFeSiB free layer, J. Magn. Magn. Mater. **304**, e276 (2006).